# Quantification of sulfated polysaccharides in urine by the Heparin Red mix-and-read fluorescence assay

Ulrich Warttinger[1], Roland Krämer[1]

Correspondence to:
Roland Krämer, phone 0049 6221 548438, fax 0049 6221 548599
E-mail: kraemer@aci.uni-heidelberg.de

1 Heidelberg University, Inorganic Chemistry Institute, Im Neuenheimer Feld 270, 69120 Heidelberg, Germany.

**Abstract**

Quantification of sulfated polysaccharides in urine samples is relevant to pharmacokinetic studies in drug development projects and to the non-invasive diagnosis and therapy monitoring of mucopolysaccharidoses. The Heparin Red Kit is a particularly simple and user friendly fluorescence assay for the quantification of sulfated polysaccharides and has recently emerged as a novel tool for the monitoring of their blood levels during pharmacokinetic studies in clinical trials. The standard protocol for the blood plasma matrix is, however, not suited for urine samples due to matrix interference. The present study identifies inorganic sulfate as the interfering component in urine. The sulfate level of urine is typically 1-2 orders of magnitude higher compared with the blood plasma level. Addition of either hydrochloric acid or magnesium chloride counteracts sulfate interference but still enables sensitive detection of sulfated polysaccharides such as heparin, heparan sulfate and dermatan sulfate at low microgramm per milliliter levels. This study extends the application range of the Heparin Red Kit by a simple modification of the assay protocol to the direct quantification of various sulfated polysaccharides in human urine.

## Introduction

Sulfated polysaccharides are complex biomacromolecules with a wide variety of biological activities. They comprise both approved drugs, such as the widely used anticoagulant heparin, and various drug candidates in preclinical or clinical development. Due to their inherent structural heterogeneity, quantification of sulfated polysaccharides in biological matrices is a challenging task. In the clinical setting, heparin blood levels are quantified by the antithrombin-mediated effect on coagulation or on specific coagulation factors. Such assays can not be readily transferred to other biological matrices. Heparin quantification in urine, for instance, is achieved by time consuming and challenging protocols that involve either radiolabeling [1, 2] or the isolation of heparin followed by colorimetry, possibly including electrophoresis [3, 4]. Depending on the analytical method, recovery of injected heparin doses in urine between 1 and 39 weight % were desribed. One study [4] has identified in urine in approximately 1:1 ratio both unchanged heparin and a partially desulfated metabolite in which the average sulfation per monosaccharide is reduced from 1.29 to 0.99. Recently there has been a renewed interest in heparin formulations for oral delivery [5, 6], and quantification of excreted heparin in urine is part of the pharmacokinetic analysis.[7] Urinary levels between 0,2 and 4 µg/mL have been reported after oral uptake of a 1000 U heparin / kg body weight dose. [7]  Urine levels of endogenous sulfated polysaccharides (glycosaminoglycans, respectively) are elevated in several types of mucopolysaccharidoses, a group of genetic disorders caused by deficiency of enzymes needed to degrade the polysaccharides. Depending on MPS type, urinary levels of heparan sulfate and dermatan sulfate, for instance, can be ten to hundreds of times higher than the normal range.[8, 9] Dye-based methods (Berry spot test and dimethylmethylene blue photometry [10]) that determine total urinnary glycosaminoglycans are routinely applied to the initial diagnostic screening for MPS. Quantification of specific polysaccharides in urine supports identification of the MPS type and therapy monitoring but requires more sophisticated analytical methodology such as HPLC-MS.[8,9]

Heparin Red is a new analytical tool for the direct detection of sulfated polysaccharides in complex biological matrices by a simple mix-and-read assay. It is a polycationic fluorescent dye (scheme 1) that forms non-fluorescent aggregates with the polyanionic target (scheme 2).[11, 12] Two assays based on Heparin Red have become commercially available [13] and been applied to the quantification of a variety of sulfated polysaccharides and biological matrices.[14-21] The Heparin Red Kit is of particular use for detections in a blood plasma matrix and has emerged as a new tool for pharmacokinetic studies of non-anticoagulant heparins that are difficult to quantify by other methods.[14, 22, 23] The standard protocol of this assay for plasma samples is, however, not suited for urine due to significant matrix interference. Another assay Heparin Red Ultra was reported to detect heparin in urine but is not sensitive to polysaccharides with a lower sulfation degree.[21]

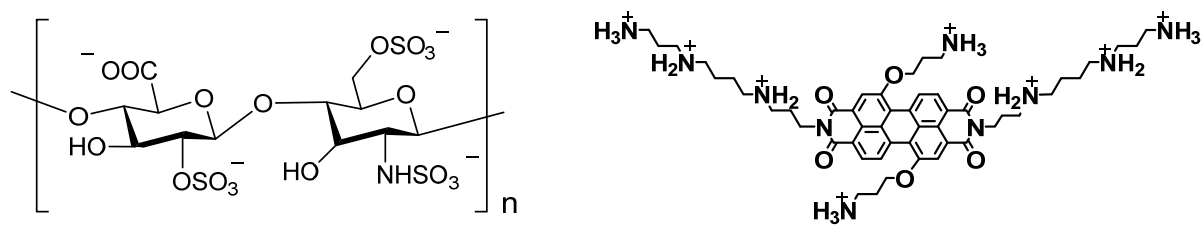

**Scheme 1**. Major repeating disaccharide units of the sulfated polysaccharide heparin (left). Structure of the polycationic fluorescent probe Heparin Red (right).

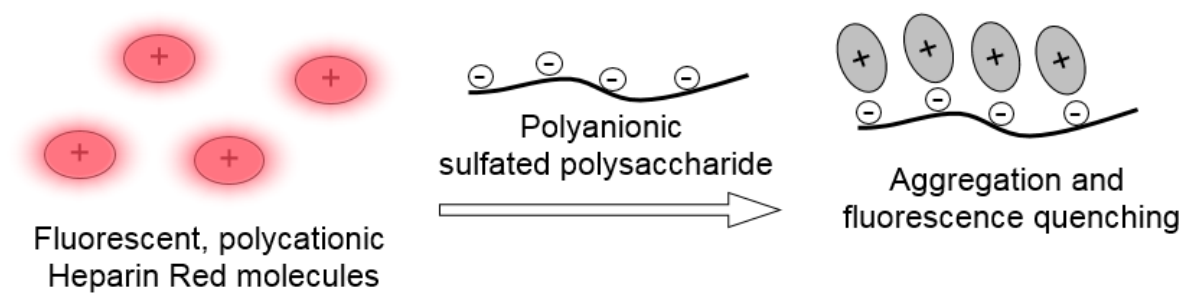

**Scheme 2.** Schematic representation of fluorescence quenching of the molecular probe Heparin Red in the presence of a polyanionic sulfated polysaccharide.

In this contribution, we identify the interference of the Heparin Red Kit in urine samples as inorganic sulfate and describe simple modifications of the assay protocol to overcome this interference. The adapted protocol enables not only the detection of heparin but also of less sulfated polysaccharides such as dermatan sulfate and heparan sulfate in urine.

## Materials and Methods

### Instrumentation

*Fluorescence measurements*
Fluorescence was measured with a microplate reader Biotek Synergy Mx (Biotek Instruments, Winooski, VT, USA), excitation at 570 nm, emission recorded at 605 nm, spectral band width 17 nm, gain 100, read height of 8 mm. Black 96 well microplates for fluorescence, polystyrene, Item No 655076, were purchased from Greiner Bio-One GmbH, Frickenhausen.

*pH measurements*

The pH value in DMSO-water 4:1 mixtures was measured with a glass electrode (InLab® Micro, SemiMicro, Mettler Toledo) that was freshly calibrated using aqueous buffers pH 4.0 and pH 7.0.

**Reagents**

*Heparin Red® Kit*

The Heparin Red Kit was a gift from Redprobes UG, Münster, Germany.

*Sulfated polysaccharides*

Unfractionated heparin sodium salt from porcine intestine mucosa was purchased from Sigma-Aldrich GmbH, Steinheim, product number H5515, Lot SLBK0235V. Heparan sulfate from porcine mucosa, highly purified fractions I and III (Cat No GAG-HS I and GAG-HS III) were purchased from Iduron Ltd, Cheshire, UK. Dermatan sulfate sodium from porcine mucosa, product number C3788 (referred to as chondroitin sulfate B), batch SLBM9912V, was purchased from Sigma-Aldrich (Taufkirchen, Germany). According to the certificate of analysis of the provider, the batch contains 9% water, 6.5% S and 8.6% Na.

*Urine*

Urine 1 is a commercially available pooled normal human urine Katalog Nr: IR100007P-1000, Lot 17317, from Innovative Research. Urine 2 is a pool urine from three healthy donors. Urines were stored at -20°C and thawed before use. Spiked urine samples were either used directly after spiking, or stored at -20°C and thawed prior to use.

*Other*

Aqueous solutions were prepared with HPLC grade water purchased from VWR, product No 23595.328. Inorganic salts were >99% pure. 18-crown-6 was purchased from Sigma Aldrich, product number 186651, Lot 0000041795. Hydrochloric acid (1.0 M), product number 35328, Lot # SZBF2050V, was purchased from Sigma-Aldrich (Taufkirchen, Germany).

**Assays**

*Heparin Red® Kit, standard protocol*

The protocol of the provider for a 96-well microplate assay with plasma samples was followed. Mixtures of Heparin Red solution and 9 mL Enhancer solution (volume ratio 1:90) were freshly prepared. Eventually (figure 1), 59,4 µL of a 1 M sodium acetate solution replaced the same volume of Enhancer solution to achieve a 6.6 mM sodium acetate concentration in the reagent mixture. 20 µL of the sample was pipetted into a microplate well, followed by 80 µL of the Heparin Red – Enhancer mixture. For sample numbers > 10, a 12-channel pipette was used for addititon of the Heparin Red – Enhancer solution. The microplate was introduced in the fluorescence reader and mixing was performed using the plate shaking function of the microplate reader (setting "high", 3 minutes). Immediately after mixing, fluorescence was recorded within 1 minute.

*Heparin Red® Kit, protocol modified fo urine samples by addition of HCl or $MgCl_2$*

The protocol of the provider for a 96-well microplate assay with plasma samples was modified as follows: To a freshly prepared mixture of Heparin Red solution and either 8.858 mL or 8.775 mL Enhancer solution (volume ratio 1:90), either 142 µL of a 1 M HCl or 225 µL of a 2 M $MgCl_2$ solution was added to achieve a 15.6 mM HCl concentration or 50 mM $MgCl_2$ concentration, respectively, in the reagent mixture. 20 µL of the sample was pipetted into a microplate well, followed by 80 µL of the reagent mixture. For sample numbers > 10, a 12-channel pipette was used for addititon of the Heparin Red – Enhancer solution. The microplate was introduced in the fluorescence reader and mixing was performed using the plate shaking function of the microplate reader (setting "high", 3 minutes). Immediately after mixing, fluorescence was recorded within 1 minute.

*Barium sulfate precipitation*

The urine sample was mixed with 9 vol% of a 100 mM $BaCl_2$ solution and kept in a 1.5 mL Eppendorf tube at ambient temperature for 10 minutes. The mixture was then centrifuged at 10000 rpm for 5 minutes.

*Sulfate assay*

Sulfate concentrations in urine were determined by the commercial Sulfate Test 101812 (Merck KGaA, Darmstadt), based on photometric turbidity measurement after precipitation as $BaSO_4$. Instructions of the provider were followed. A 100-fold dilution of the urine sample was applied.

## Results and discussion

**Sulfate interferes with the Heparin Red assay**

A preliminary screening of the abundant inorganic ions in urine (table 1) revealed that only sulfate ($SO_4^{2-}$) quenches Heparin Red fluorescence significantly. From a refined analysis of sulfate interference, we conclude that not only the sulfate concentration but also the pH of the assay mixture strongly influences fluorescence quenching. The assay medium is a mixture of about 76 vol% DMSO, 4 vol% acetic acid and 20 vol.% aqueous sample. There are several literature reports [24-27] about reproducible pH measerumnts with a glass electrode in a DMSO water 80:20 mixture. We have therefore applied this technique to pH determination of the assay mixtures.

| Compound | Reference range for urine [28] * |
|---|---|
| $Na^+$ | 27-167 mM |
| $Mg^{2+}$ | 2.5-8.5 mM |
| $Ca^{2+}$ | 2.5-7.5 mM |
| $K^+$ | 23-67 mM |
| $Cl^-$ | 67-167 mM |
| $H_2PO_4^-/HPO_4^{2-}$ | 9-28 mM ** |
| $SO_4^{2-}$ | 12 mM (5-31 mM) *** |

**Table 1**. Reference ranges of the major inorganic ions present in human urine (usually defined as the set of values 95% of the normal population falls within). * Concentration in urine derived from values for excretion over 24 hours [28], assuming a daily urine volume of 1.5 L [29]. ** This concentration value refers to the sum of $H_2PO_4^-$ and $HPO_4^{2-}$ since the pKa of $H_2PO_4^-$ is 6.8 and both species may be present within the reference pH range 4.5-8 [28] of urine. *** Sulfate determination is not part of routine urinalysis. 12 mM is the average sulfate concentration in normal urine from 110 healthy donors, by ion chromatography.[30] 5-31 mM relates to the reference range given by several clinical laboratories.[31, 32] Sulfate excretion is highly dependent on dietary protein uptake.[33]

Figure 1 displays the effect of sulfate concentration on Heparin Red fluorescence. The pH of the assay mixture was kept approximately constant at 6.0 by adding 6.6 mM sodium acetate to the reagent solution, thus creating an "acetate buffer" (although at a large excess of acetic acid). This mimics the effect of basic urinary buffer components such as $HPO_4^{2-}/H_2PO_4^-$ that are expected to react with acetic acid and form acetate upon mixing the reagent solution with the urine sample. It is obvious from the high fluorescence in the absence of sulfate that the acetate anion itself does not interfere with the assay. When the assay mixture is prepared with a sample of urine 1, the resulting pH 6.0 matches that of the acetate buffered medium applied for the detections in figure 1.

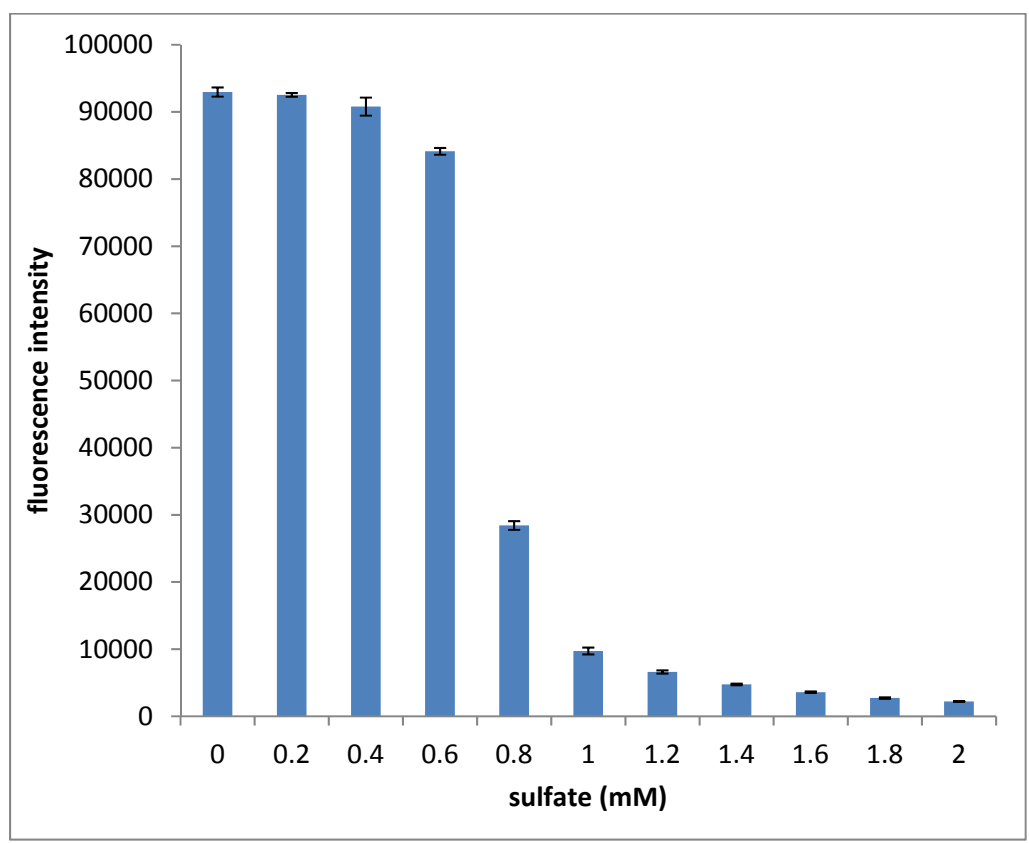

**Figure 1.** Fluorescence response of Heparin Red to aqueous samples containing $Na_2SO_4$ in varying concentration. The standard protocol for the Heparin Red Kit was modified by adding 6.6 mM sodium acetate to the Enhancer solution (see Materials and Methods). Manually performed microplate assay. Excitation at 570 nm, fluorescence emission at 605 nm. Averages of duplicate determinations. The pH of the acetate-buffered assay solution at c(sulfate) = 0 mM was 6.0. The pH of the acetate-buffered assay solution at c(sulfate) = 2 mM was 6.0.

The effect of pH variation of the assay solution at constant sulfate concentration of 2 mM is shown in figure 2. If a sample containing 2 mM $KHSO_4$ is mixed with the reagent solution, pH 4.8 is measured in the assay solution and no fluorescence quenching of Heparin Red observed. The pH of this mixture is similar to that in the absence of $KHSO_4$, suggesting that $HSO_4^-$ dissociates only to a minor extent into protons and $SO_4^{2-}$. These observations support the idea that the $HSO_4^-$ monoanion is not interfering with the assay. The pH of the assay solution was gradually increased by adding NaOH to the $KHSO_4$ sample prior to mixing with the reagent solution. This went along with a decrease of dye fluorescence due to conversion of $HSO_4^-$ to $SO_4^{2-}$. A mixture of 2 mM $KHSO_4$ and 10 mM NaOH is expected to not only generate 2 mM $SO_4^{2-}$ in the sample but also an acetate/acetic acid buffer medium in the assay solution comparable to that described for figure 1. This is confirmed by measuring pH 6.0, identical to that of the assay solution of figure 1 at 2 mM sulfate.

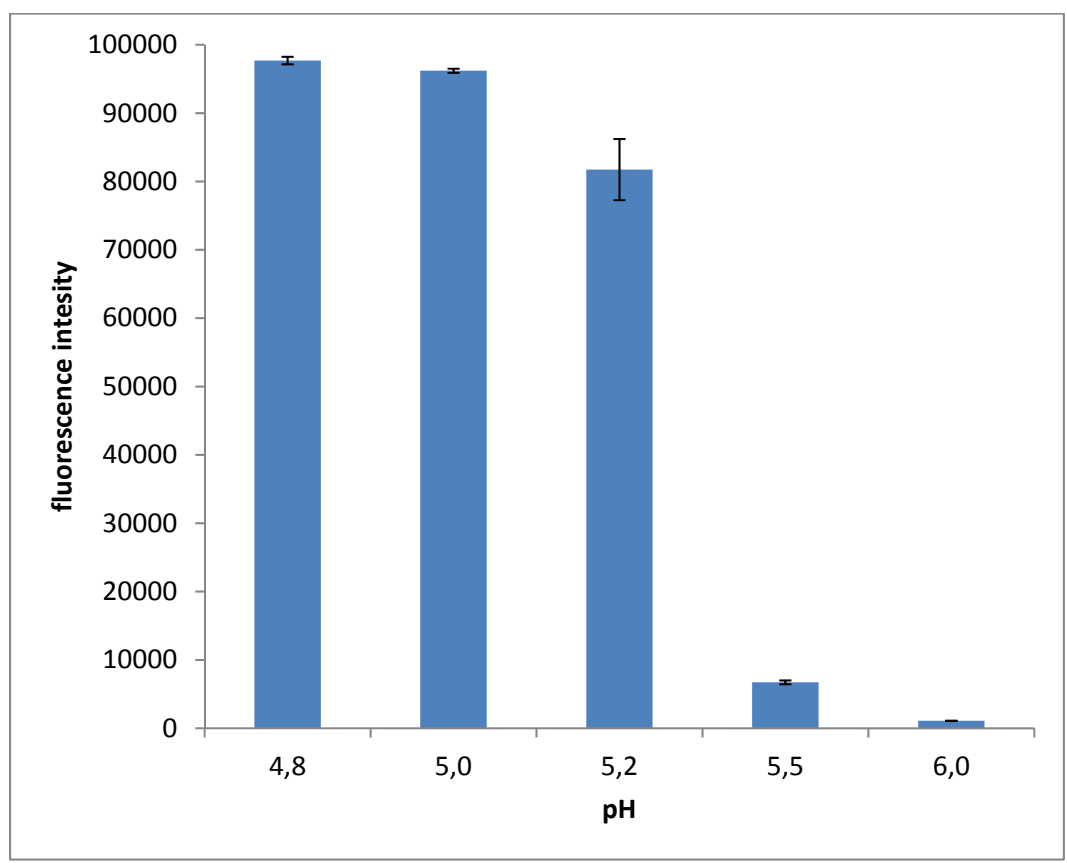

**Figure 2.** Fluorescence response of Heparin Red to aqueous samples containing 2 mM sulfate at varying pH of the assay solution (obtained by mixing the Enhancer-Heparin Red reagent solution with the sample). The standard protocol for the Heparin Red Kit (see Materials and Methods) was applied. Variation of pH was achieved by adding varying concentrations of NaOH (0 mM at pH 4.8 to 10 mM at pH 6.0) to an aqueous $KHSO_4$ sample. Manually performed microplate assay. Excitation at 570 nm, fluorescence emission at 605 nm. Averages of duplicate determinations.

Note that the pKa values of acids that form anionic conjugated bases are shifted to higher values in DMSO:water 80:20 relative to aqueous solution due to less effective anion solvation by the organic solvent . The pKa of acetic acid in DMSO:water 80:20 is 8.0 [25, 34], vs 4.8 in aqueous solution. For $HSO_4^-$, we also expect a significant pKa increase relative to the value 1.9 for aqueous solution. We measured pH 6.7 for a 1:1 mixture of $KHSO_4$ and $Na_2SO_4$ (5 mM : 5 mM) in DMSO:water 80:20, suggesting that the value corresponds to the pKa of $HSO_4^-$ in this medium.

**Sulfate is the interfering component in urine**

Fig. 3 compares the fluorescence of Heparin Red with human plasma, phosphate buffered saline (PBS) and urine 1 samples, following the standard assay protocol for the Heparin Red Kit. Only the urine matrix triggers a strong fluorescence response of the dye so that this matrix is not well suited for the detection of sulfated polysaccharides.

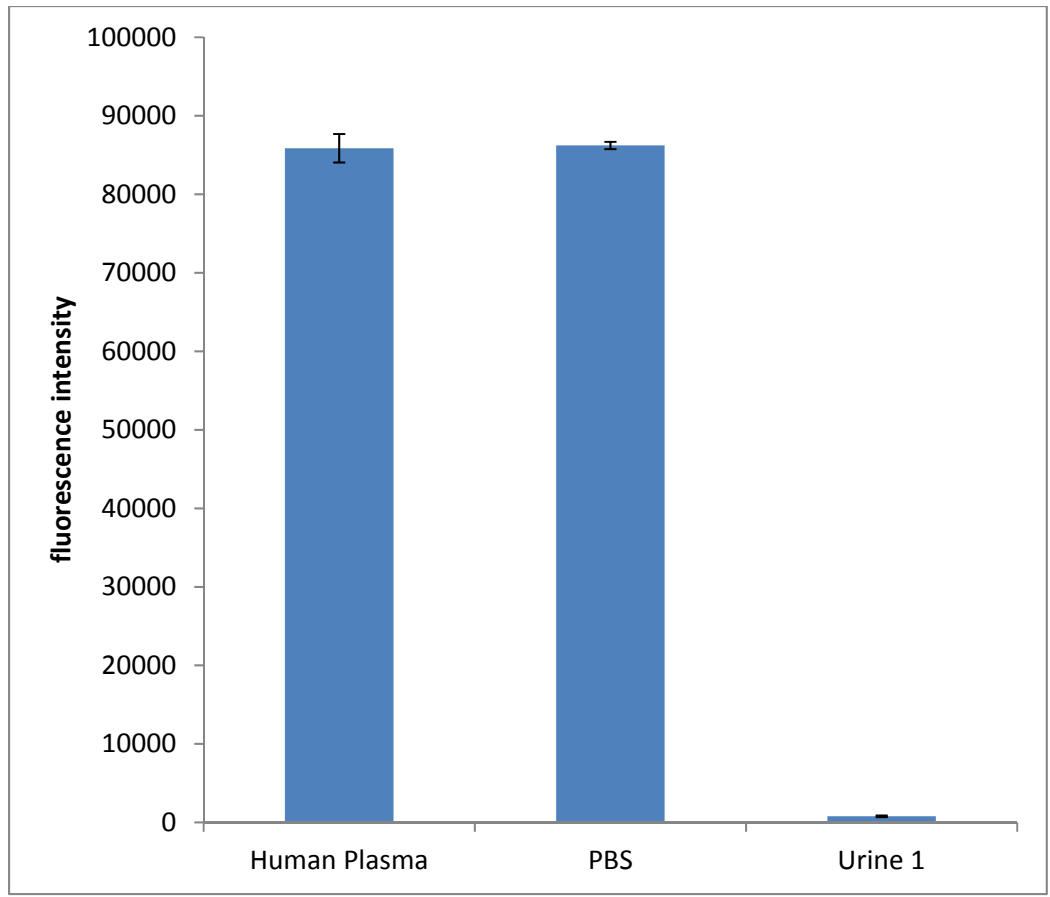

**Figure 3.** Fluorescence response of Heparin Red to a human plasma, a phosphate buffered saline (PBS) and a urine 1 sample, following the standard protocol for the Heparin Red Kit (see Materials and Methods). Manually performed microplate assay. Excitation at 570 nm, fluorescence emission at 605 nm. Averages of duplicate determinations.

Sulfate in the urine 1 sample was quantified by a commercial assay that detected 4.7 mM, a concentration at the lower limit of the reference range for urine (table 1). Strong fluorescence quenching at this sulfate concentration is in line with the data in figure 1. The urine level of sulfate (table 1) is 1-2 orders of magnitude higher than the plasma or serum level which is normally in the range 0.25 to 0.4 mM [35-37]. This readily explains why sulfate interference is not relevant to plasma samples since this concentration does not trigger fluorescence quenching of Heparin Red (figure 1, figure 3).

Sulfate interference in urine is further substantiated by its removal by precipitation as $BaSO_4$, through addition of 10 mM barium chloride to the urine sample and removal of the precipitate by centrifugation. After this pretreatment, the urine sample does no longer quench the fluorescence of Heparin Red, see figure 4 B. It is unlikely that fluorescence recovery is triggered by excess barium chloride (rather than sulfate) since subsequent masking of $Ba^{2+}$ by the macrocyclic ligand 18-crown-6 (100 mM) does not reduce fluorescence intensity (figure 4 C). Moreover, if 18-crown-6 is added to the urine sample prior to $BaCl_2$, $BaSO_4$ precipitation is largely prevented by complexation of the metal ion, resulting in significant interference by sulfate that has not been removed from the sample (figure 4 D). An additional control experiment using 20 mM NaCl confirmed that the chloride of the 10 mM $BaCl_2$ additive has no effect on fluorescence response to the urine sample (data not shown).

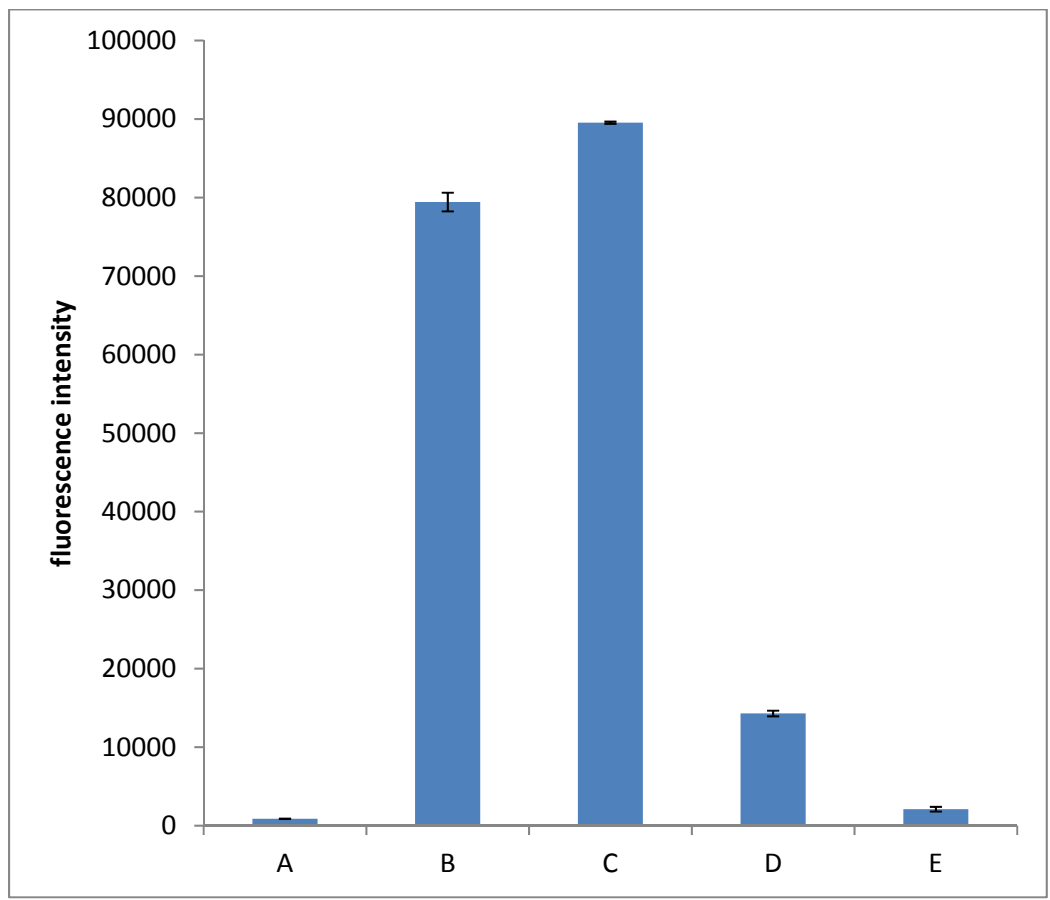

**Figure 4.** Fluorescence response of Heparin Red to A) urine 1; B) urine 1 + 10 mM $BaCl_2$, after removal of precipitate; C) solution B, removal of precipitate + 100 mM 18-crown-6; D) urine 1 + 100 mM 18-crown-6 + 10 mM $BaCl_2$, very minor precipitate; E) urine 1 + 100 mM 18-crown-6. Standard protocol for the Heparin Red Kit was followed (see Materials and Methods). Manually performed microplate assay. Excitation at 570 nm, fluorescence emission at 605 nm. Averages of duplicate determinations.

**Sulfate interference is overcome by modifiying the Heparin Red assay protocol**

Considering the strong pH dependence of sulfate interference of the Heparin Red assay (figure 2), we established addition of hydrochloric acid to the assay solution as a strategy to overcome sulfate interference and enable the detection of sulfated polysaccharides in urine samples. The monoanionic sulfate ester moieties in sulfated polysaccharides are much less basic than the $SO_4^{2-}$ dianion. We therefore anticipated selective protonation of $SO_4^{2-}$ to $HSO_4^-$ without affecting the sulfate ester groups by proper control of the pH of the assay solution. In practice, the standard assay protocol was modified simply by adding 15.6 mM HCl to the reagent mixture, prior to its application to the sample (see Materials and Methods for details). The presence of HCl effected a much higher fluorescence with urine samples, comparable to that of aqueous or PBS samples. With the modified protocol, not only heparin, having an average charge density of -1.7 per monosaccharide and a sulfation degree of 1.2 per monosaccharide, but also structurally related polysaccharides with a lower sulfation degree (table 2) such as heparan sulfate and dermatan sulfate can be recovered as spikes in urine samples (figure 5). Moreover, the response curves display good linearity within the concentration range 0-10 µg/mL polysaccharide.

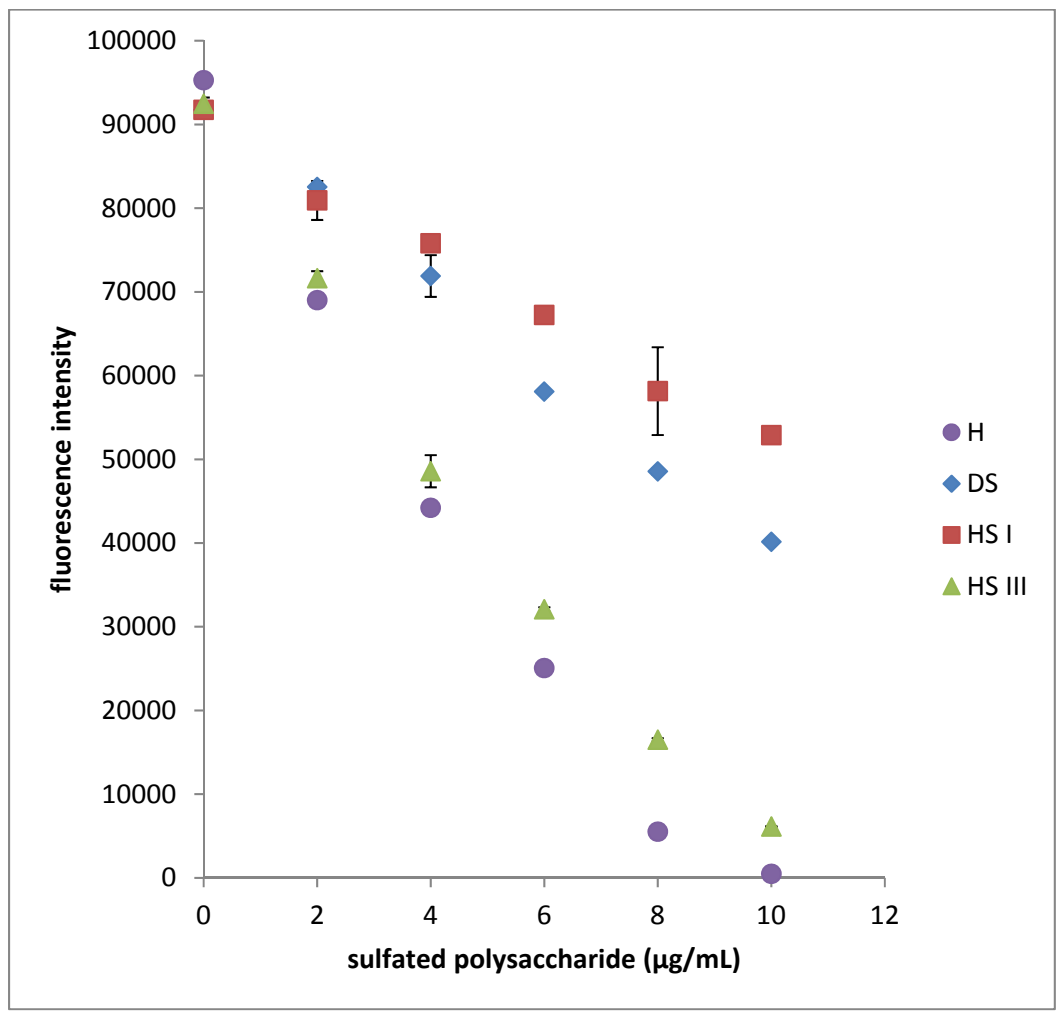

**Figure 5.** Fluorescence response of Heparin Red to samples of urine 1 spiked with heparin (H), heparan sulfate fraction I (HS I), heparan sulfate fraction III (HS III), and dermatan sulfate (DS) in varying concentrations. A modifed protocol for the Heparin Red Kit including addition of 15.6 mM HCl to the reagent mixture was followed (see Materials and Methods for details). Manually performed microplate assay. Excitation at 570 nm, fluorescence emission at 605 nm. Averages of duplicate determinations.

The pH of the HCl containing assay solution after mixing with the urine sample was 3.1, and in line with the results outlined in figure 2 an interference by urinary sulfate is not expected under these conditions due to its conversion to non-interfering $HSO_4^-$.

The sensitivity of the assay for a specific sulfated polysaccharide depends on the sulfation degree of the latter (table 2). The higher the sulfation, the less polysaccharide is needed to form a charge neutral complex with Heparin Red [12], and the more sensitive the assay is. Considering pKa 8 for acetic acid in the DMSO water 80:20 assay solution [25, 34], we assume protonation of the carboxylate groups of the analytes (scheme 1). Therefore, only the sulfate groups remain for electrostatic binding to polycationic Heparin Red (scheme 1, 2).

| Sulfated polysaccharide | Average sulfation per monosaccharide | Average charge density per monosachharide | Ref |
|---|---|---|---|
| H (heparin) | 1.2 | -1.7 | [38] |
| DS (dermatan sulfate) | 0.55 | -1.05 | [18] |
| HS I (heparan sulfate fraction I) | 0.38 | -0.88 | [15] |
| HS III (heparan sulfate fraction III) | 0.88 | -1.38 | [15] |

Table 2. Average sulfation and charge (including carboxylate groups) per monosaccharide of sulfated polysaccharides tested with the Heparin Red assay in figure 5.

Different pool urine samples (urine 1, urine 2) show very similar response curves, as exemplified by detection of a HS III spike (figure 6). Moreover, increasing the HCl concentration in the reagent mixture from 15.6 to 30 mM does not significantly change the response curve for HS III (data not shown). Apparently, 15.6 mM HCl both compensates the buffer capacity and protonates interfering sulfate in the urine samples used in this study.

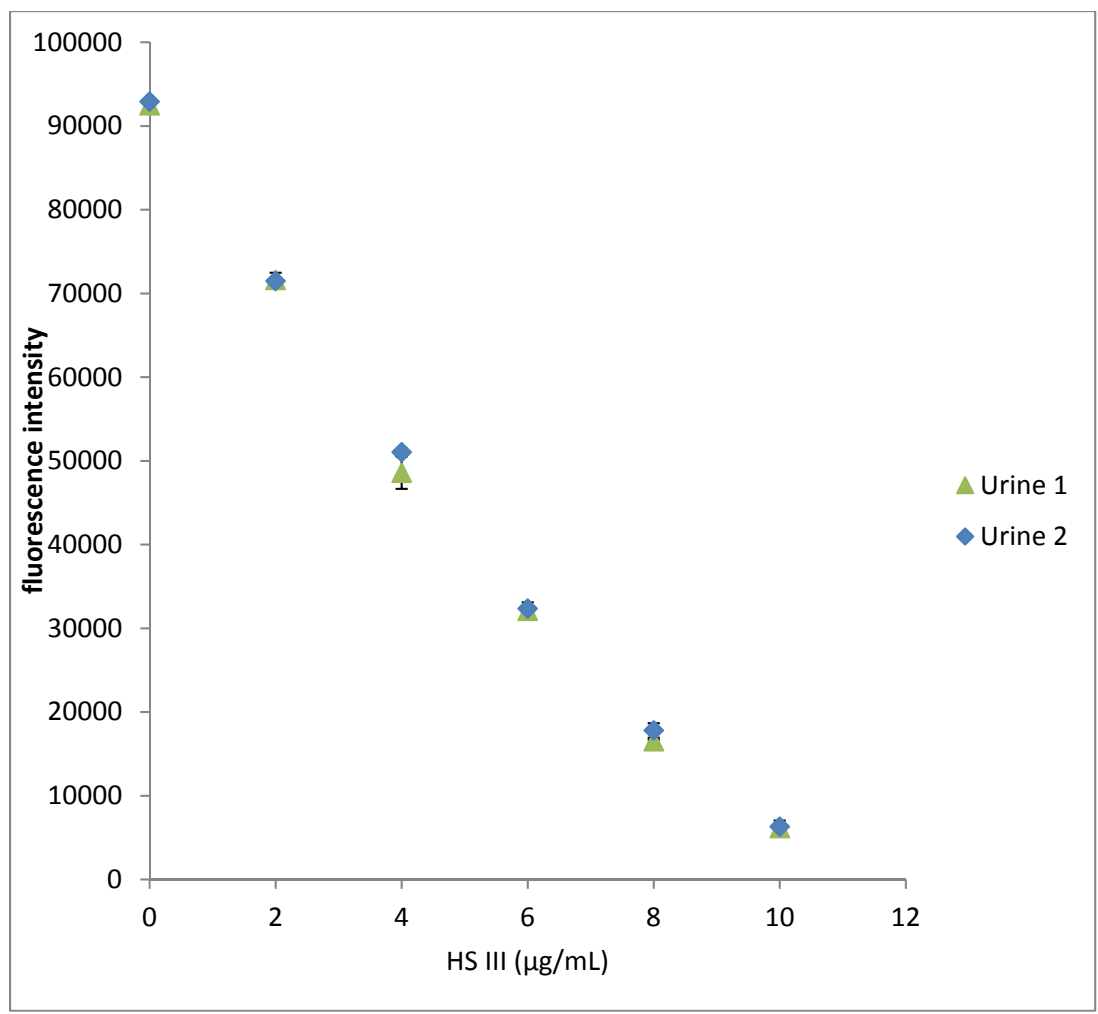

**Figure 6.** Fluorescence response of Heparin Red to different pool urine samples, urine 1 and urine 2, spiked with HS III. A modifed protocol for the Heparin Red Kit including addition of 15.6 mM HCl to the reagent mixture was followed (see Materials and Methods for details). Manually performed microplate assay. Excitation at 570 nm, fluorescence emission at 605 nm. Averages of duplicate determinations.

We have previously reported the detection of heparin in urine by Heparin Red Ultra, another commercially available assay.[21] Heparin Red Ultra does, however, not respond to less sulfated polysaccharides such as dermatan sulfate. Human serum albumin (HSA) that may be present in trace amounts in urine interferes with the assay. We have interpreted this by conversion of HSA to a polycationic strong heparin binder in the acidic reaction medium of the assay, masking the response of Heparin Red.[21] HSA spikes have a comparable masking effect if the modified Heparin Red Kit protocol with addition of HCl is applied (figure 8).

We have therefore evaluated another protocol modification, the addition of 50 mM $MgCl_2$ to the reagent solution [20], for detection of sulfated polysaccharides in urine samples. $MgCl_2$ also overcomes the strong fluorescence quenching by urinary sulfate and enables the sensitive detection of the polysacchairdes listed in table 2 (figure 7). The pH 4.5 of the 50 mM $MgCl_2$ containing reagent solution after mixing with water (80:20) is the same as in the absence of $MgCl_2$, suggesting that masking of sulfate is not a pH effect but due to direct binding of the lewis-acidic diavalent magnesium ion to sulfate. In contrast to the HCl containing medium, detections in the $MgCl_2$ medium are much less affected by trace amounts of HSA (figure 8). In the general population, the average urinary HSA concentration is 5 µg/mL, and less than 25% have urinary HSA levels > 15 µg/mL (the values were derived from mg HSA / g creatinine data [39], asuming an average urinary creatinine concentration in the general population of 1,3 g / L [40]). HSA in urine is commonly determined by immunoassays and measured concentrations must not essentially correlate with the effect of the HSA spikes in figure 8 (using a commercial HSA sample isolated from human serum) since different molecular forms of albumin may be present in urine.[41]

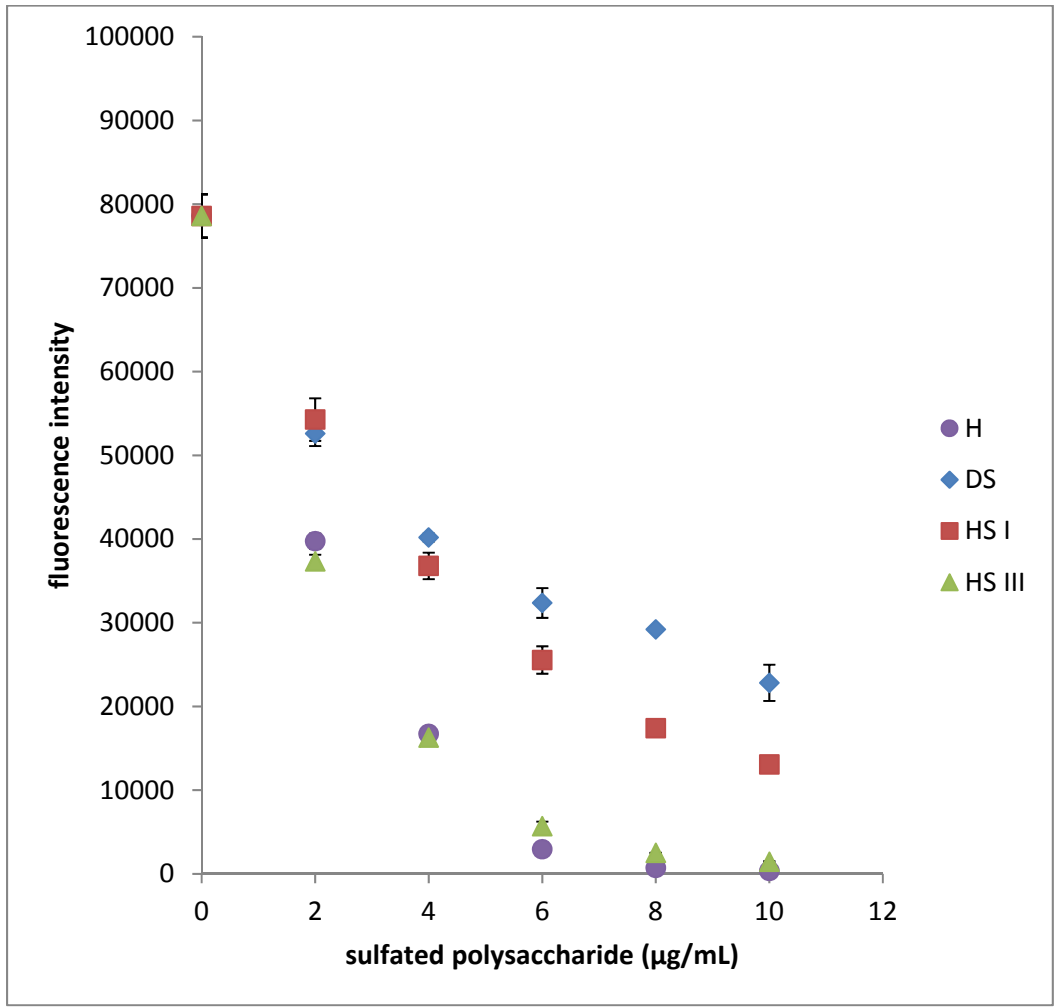

**Figure 7.** Fluorescence response of Heparin Red to samples of urine 1 spiked with heparin (H), heparan sulfate fraction I (HS I), heparan sulfate fraction III (HS III), and dermatan sulfate (DS) in varying concentrations. A modifed protocol for the Heparin Red Kit including addition of 50 mM $MgCl_2$ to the reagent mixture was followed (see Materials and Methods for details). Manually performed microplate assay. Excitation at 570 nm, fluorescence emission at 605 nm. Averages of duplicate determinations.

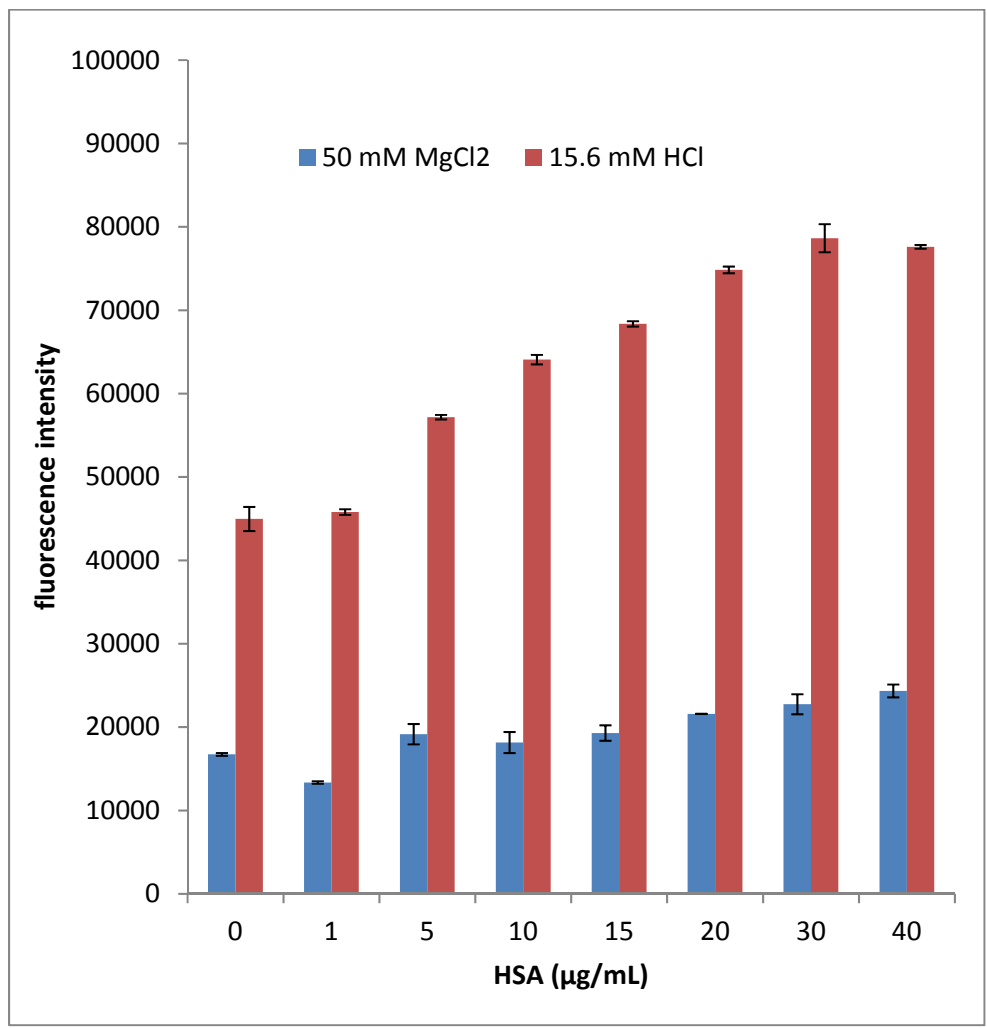

**Figure 8.** Effect of human serum albumin (HSA) spikes in varying concentration on the fluorescence response of Heparin Red to samples of urine 1 containing 4 µg/mL heparin. Modfied protocols for the Heparin Red Kit including either addition of 50 mM $MgCl_2$ or addition of 15.6 mM HCl to the reagent mixture was followed (see Materials and Methods for details). Manually performed microplate assay. Excitation at 570 nm, fluorescence emission at 605 nm. Averages of duplicate determinations.

The presence of endogenous sulfated polysaccharides in urine should be considered as a potential cause of background signal in the Heparin Red assay. According to a recent LCMS analysis [42], the overall sulfated polysaccharide concentration in human urine averages to about 4.8 µg/mL and is a mixture of 83 % chondroitin sulfates (CS, including dermatan sulfate) and 17 % heparan sulfate. Individual levels are, however, highly variable and can range from <1 µg/mL to 10 µg/mL. Average sulfation per monosaccharide is 0.32 for urinary CS and 0.35 for urinary HS (derived from the data in [42]). This corresponds to a lower sulfation compared with any sulfated polysaccharide investigated in the present study (table 2), and endogenous urinary CS and HS might be relatively poor targets for the Heparin Red assay. Fluorescence of the unspiked pool urine samples in figure 6 is comparable to that of a phosphate buffered saline (PBS) sample, suggesting that these samples do not contain detectable levels of endogenous sulfated polysaccharides. It should nevertheless be considered that even normal urine may contain CS and HS levels that generate a background signal.

HS and DS are urinary biomarkers for most types of mucopolysaccharidoses (MPS). In MPS type I-III, 100-200 times higher than normal urinary HS levels were determined by HPLC analysis.[8] A 6-10fold elevation of DS was reported for children with MPS type I and VI.[9] The modified Heparin Red assay as described in Figure 5 and 8 should readily detect such elevated overall concentrations of sulfated polysaccharides in the urine of MPS patients. The diagnostic value would be stronger if concentrations of individual sulfated polysaccharides can be determined. We plan therefore to refine the method by implementing selective enzymatic digestion of one polysaccharide in HS/DS mixtures.

**Conclusion**

Quantification of sulfated polysaccharides in urine samples is relevant to pharmacokinetic studies in drug development projects and to the non-invasive diagnosis and therapy monitoring of mucopolysaccharidoses. The Heparin Red fluorescence assay is a particular simple and user friendly analytical method for sulfated polysaccharides quantification and has recently emerged as a novel tool for the monitoring of their blood levels during pharmacokinetic studies in clinical trials. The standard protocol for the blood plasma matrix is, however, not suited for urine samples due to matrix interference. In this study we have identified inorganic sulfate ($SO_4^{2-}$) as the interfering component in urine. The sulfate level of urine is usually 1-2 orders of magnitude higher compared to the blood plasma level. Lowering the pH of the assay solution by addition of hydrochloric acid counteracts sulfate interference due to protonation of dianionic sulfate to non-interfering $HSO_4^-$ monoanion, but still enables sensitive detection of sulfated polysaccharides such as heparin, heparan sulfate and dermatan sulfate. Alternatively, addition of magnesium chloride to the assay solution overcomes sulfate interference and offers the additional advantage of largely supressing an interference by albumin that may be present in trace amount in urine samples. This study extends the application range of Heparin Red by simple modifications of the assay protocol to the direct quantification of various sulfated polysaccharides at low µg/mL levels in human urine.

**Conflict of interest.** R. Krämer holds shares in Redprobes UG, Münster, Germany. Other authors: No conflict of interest.